
\magnification 1200
\font\titlefont=cmb10 scaled\magstep1
\baselineskip 15pt
\hsize = 6.0 true in
\vsize = 8.5 true in
\hoffset 0.2 true in
\voffset 0.2 true in
\parindent=16pt
\parskip=0pt 

\def\frac#1#2{{#1\over#2}} 

\def\section#1{\vskip4ex plus4ex minus1ex
       \vbox{\centerline{\bf #1}\vskip2ex}\nobreak}

\def\date{ 
   \count0 = \year
   \advance \count0 by -1900
   \number\month/\number\day/\number\count0}

\def\Ref#1{\thinspace[\thinspace #1\thinspace]}

\def\GeVc2{\hbox{GeV$\!/c^2$}}
\def\pb{\hbox{pb}}

\nopagenumbers
\footline={\ifnum\pageno>0 \hfil\folio\hfil \else \hfil \fi}

\pageno=0
\null\vskip -21pt
\rightline{PVAM-HEP-93-1}
\rightline{SSCL- PP- 189}
\rightline{UTAPHY-HEP-7}
\rightline{\date}
\vskip 2ex
\centerline{\titlefont Higgs particle detection using jets}
\vskip 6ex
\centerline{T. Garavaglia,$^{(1)}$ Waikwok Kwong,$^{(2)}$
and Dan-Di Wu$^{(1,3)}$}
\vskip 2ex
\centerline{\it $^{(1)}$Superconducting Super Collider Laboratory,{\rm *}
   Dallas, Texas 75237.}
\centerline{\it $^{(2)}$Dept.~of Physics, University of Texas at Arlington,
   Arlington, Texas 76019.}
\centerline{\it $^{(3)}$High Energy Physics,{\rm **} P. O. Box 355, Prairie
   View A \& M University,}
\centerline{\it Prairie View, Texas 77446.}

\vskip42pt
\vbox{ \leftskip11pt \rightskip15pt
We study the possibility of detecting the Higgs boson in the intermediate
mass range via its two jet channel. We consider only Higgs bosons produced
in association with a $t \bar{t}$ pair. Both $t$ and $\bar{t}$ are required
to decay  semileptonically to reduce the QCD background. The signal is
compared with the main background, $t \bar{t} + 2$ jets, after appropriate
cuts.  A sizable signal above background is
seen in our simulation at the parton level. Use of the $t\bar{t}Z$ channel with
$Z$ decaying to $l^+ l^-$ is
suggested for eliminating theoretical uncertainties in determining the
$t \bar{t}H$ signal.

\vskip 30pt
PACS numbers: 14.80.Gt., 13.85.Qk
} 

\vfil

\hrule width 1in \vskip1ex
\item{*} Operated by the Universities Research Association, Inc., for the
      U.~S.~Department of Energy under contract DE-AC35-89ER40486.
\item{**} Supported in part by the Texas National Research
      Laboratory Commission  through the Particle Detector Research Center
						under contract RCFY 9208.
\eject

\section{INTRODUCTION}

The Standard Model Higgs with mass up to 85 \GeVc2\ could be discovered at
the planned LEP 200 experiments. At the SSC or the LHC on the other hand it
should not be difficult to find a Higgs boson with mass between 140--800
\GeVc2, which can decay to two $Z$ bosons, one of which may be virtual.
To cover the mass gap between 85 and 140 \GeVc2, the rare decay $H \to
\gamma\gamma$ was suggested\Ref{1--3}, where the Higgs boson $H$ is
produced in association with either a $W^\pm$ boson or a $t \bar{t}$ pair
in $pp$ scattering experiments.  A charged hard lepton from $W^\pm$ or one
of the $t$ quarks is required as a trigger in order to reduce the QCD
background. However, since the $2\gamma$ mode is a rare decay, it may be
difficult to extract this signal from the complicated background\Ref{4}.

We study instead the possibility of searching for an intermediate Higgs
boson in the two-jet channel by reconstructing its invariant mass. We
consider Higgs bosons that are produced in association with $t \bar{t}$
pairs only.  This allows both the $t$ and $\bar{t}$ to be tagged using
semileptonic decays to cut down on the QCD background.  We are motivated by
recent works on jet spectroscopy. The GEM Collaboration\Ref{5} has
presented mass plots where $W$ and $t$ peaks were successfully
reconstructed from two and three jets in
computer generated multi-jet events. Similar method
are proposed for charged Higgs searches at HERA\Ref{6} and for top searches
at the TEVATRON\Ref{7}. For earlier studies of Higgs searches using jets
see Ref.\Ref{8}.

\section{SIGNAL AND BACKGROUND}

The main process of interest is
$$
  pp\to t\bar{t} H + X,  \eqno(1)
$$
with
$$\eqalignno{
  t       &\to b + (W^+\to l^+ +\nu),                &(2) \cr
  \bar{t} &\to \bar{b} + (W^-\to l^- +\bar{\nu}),    &(3) \cr
  H       &\to b\bar{b} \quad{\rm or}\quad c\bar{c}, &(4) \cr}
$$
where $l$ stands for either an electron or a muon. The final signal we are
looking for is
$$
  {\rm 2~leptons + 4~jets + missing}~E_T.  \eqno(5)
$$
Dicus and Willenbrock\Ref{1} have shown that $t\bar{t}H$ production at $pp$
colliders can be very well approximated by two-gluon fusion alone. The
tree level cross section as a function of the Higgs mass can be found in
Kunzst et al.\Ref{1} for $m_t = 140~\GeVc2$. It varies from 12 to 3 pb as
the Higgs mass changes from 80 to 140 \GeVc2. As an example, we take $m_H
= 100~\GeVc2$, which gives a cross section of 8 pb; our signal (5) has a
combined cross section $\times$ branching ratio of
$$
  \sigma(pp\to t\bar{t}H) B^2(t\to bl^+\nu) B(H\to q\,\bar{q})
  = 8~\pb \times (2/9)^2 \times 0.97 = 0.4~\pb
\eqno(6)
$$
at the SSC. This is to be compared with 6 fb for $H \to \gamma\gamma$ with
a single lepton tag.

To reduce most of the QCD background, we put a 20 GeV cut on the minimum
$p_T$ of each of the leptons and jets and place a 40 GeV minimum on the
total missing $E_T$. An isolation cut is then applied to each lepton,
requiring it to be at least $\sqrt{\Delta\eta^2 + \Delta\phi^2} = 0.4$
units away from any of the four jets. This has the effect of guaranteeing
that almost all remaining events are of the type $t\bar{t} + 2$ hard jets.
The same amount of separation is also applied to $e^+e^-$ and $\mu^+\mu^-$
pairs to avoid QED background.

We now look at the possible sources of background to our signal. There are
three major types of background events:

(1) Sources of background other than $t\bar t$ that produce $l^+ l^-$ pairs
are suppressed by at least two factors of either $\alpha_{weak}$ or
$\alpha_{em}$, except when the $e^+e^-$ or $\mu^+\mu^-$ pairs are on the
$Z$ mass-shell. Furthermore, these events do not have intrinsic missing
$E_T$ and will probably not pass the missing $E_T$ cut. Thus, non
$t\bar{t}$ contributions to the background are likely to be small.

(2) By far, the most important background is the QCD process
$$
  pp \to t{\bar t} + 2~\rm jets,
\eqno(7)
$$
where the jets come mostly from initial state radiation. Both the $t$ and
$\bar{t}$ decay, as before, semileptonically to give the required
signature.

The inclusive cross section for $pp \to t\bar{t}$ has so far been calculated
to order $\alpha_s^3$ only\Ref{8a}.  The cross section at the SSC was
estimated by Baer et al.\Ref{10} to be 16 nb, while Barger et al.\Ref{11}
gave 10 nb. To be conservative, we take the larger number of 16 nb for
background calculations.

Although works on higher order corrections are still in progress\Ref{9},
experience shows that $t\bar{t}$ pairs produced at the SSC are almost always
accompanied by extra high $p_T$ jets.  Here, we will make the ansatz that
there will be an average of 2.0 jets with $p_T > 20$ GeV in every $t\bar{t}$
event. By assuming a Poisson distribution for the number of extra jets, the
$t\bar{t}$ cross section with two and only two accompanying jets with $p_T >
20$ GeV is found to be 4.3 nb.  Folding in the leptonic branching ratio for
top quarks reduces this number to 0.2 nb.  This is a worst case scenario
since we do need two extra jets in the final state and the Poisson
distribution peaks at the average value.  We therefore believe this to be a
fairly conservative estimate unless the inclusive $t\bar{t}$ cross section
turns out to be significantly different from 16 nb.

There is also a combinatorial factor for forming two-jet pairs from the four
jets in each event; this enhances the background by a factor of 6. Compared
to 0.4 pb for the signal, the background is larger by a factor of $3 \times
10^3$.  We shall see later that this will be suppressed severly by
additional cuts (see Table I).  In addition, the background events will not
have any special feature in the two-jet invariant mass spectrum while the
signal will show a prominent peak around the mass of the Higgs boson.

(3) In addition to $t\bar{t}H$, there are also $t\bar{t}W^{\pm}$
and $t\bar{t}Z$  events in which the $W^{\pm}$ and $Z$ decay to two jets,
thus producing the same signature. The production cross section for
$W^{\pm}$ is about one-tenth of that for the $Z$\Ref{12}.  This together
with a smaller mass of 80 GeV makes the $W^{\pm}$ events less important. On
the other hand,  $t\bar{t}Z$ is an exact analogue of $t\bar{t}H$ for $m_H$
close to the $Z$ mass, and we expect the two to have very similar cross
sections. Baer et al.\Ref{10} and Barger et al.\Ref{11} have estimated the
cross section for $t\bar{t}Z$ to be roughly 10--12 pb for $m_t=140$. The
mass resolution for $Z\to 2$ jets at the SSC was estimated by the GEM
collaboration to be roughly ${}\pm 5$ GeV in a similar energy dependent
situation. This overlaps significantly with Higgs bosons of mass up to 100
GeV/$c^2$.

This background not only turns out to be benign, it actually works to our
advantage. Unlike the Higgs, the $Z$ also decays to charged lepton pairs
with a large branching ratio (6\%). In our case, this gives a signature of
4 leptons and 2 jets, which has very little background if we further
require two of the leptons to reconstruct to a $Z$. Therefore, the
$t\bar{t}Z$ cross section can be measured by reconstructing the two-lepton
invariant mass in  the 4 leptons + 2 jets events.  We expect that most of
the theoretical uncertainty in the ratio of the two cross sections, $pp
\rightarrow t\bar{t} H$ and $pp \rightarrow t \bar t Z$,  cancels, and the
cross section for $t\bar{t}H$ can then be reliably inferred from those of
the $t\bar{t}Z$.  Even if the Higgs and the $Z$ peaks overlap, we will
still be able to tell how much of the 2-jet invariant mass peak is due to
the $Z$ and how much is from the Higgs for a given $m_H$. In other words,
we will be able to tell whether there is a Higgs boson at a certain mass
without having to know exactly its production cross section and, if a
signal is seen, whether it is from a standard or nonstandard Higgs.
Finally, this calibration from the $Z$ is absent in the $\gamma\gamma$
channel because $Z \to \gamma\gamma$ is forbidden on account of anomaly
cancellation.

\section{SIMULATION}

Both the signal and background are simulated using Monte Carlo methods at
the parton level. As a preliminary study, simplified distributions in phase
space are used for each scattering and decay process. The total cross
sections are then normalized to published values.

For our signal, the parton cross section is given on purely dimensional
ground by
$$
  d\sigma(gg \to t\bar{t}H) \propto \frac{1}{~\hat{s}^2}\,d\Phi_3~,
\eqno(8)
$$
where $d\Phi_n$ is the $n$-body Lorentz invariant phase space:
$$
  d\Phi_n = \delta^{(4)}(P - {\textstyle\sum} p_i)\,
	    \prod_{i=1}^n \frac{d^3p_i}{2E_i}~,
\eqno(9)
$$
where $\hat{s}$ is the center of mass energy of the partons and $P$ is
their total momentum.  Each of $t,\ \bar{t}$ and $H$ is then decayed
independently to $b\,l^+\nu$, $\bar{b}\,l^-\bar{\nu}$ and $b \bar{b}$
respectively to give us the signature of 4 jets + 2 leptons + missing
$E_T$. We take $m_t = 140,\ m_H = 100~\GeVc2$ and 8 pb for the total cross
section.

For the background, we first take
$$
  d\sigma(gg \to t\bar{t}gg) \propto \frac{1}{~\hat{s}^3}\,d\Phi_4~,
\eqno(10)
$$
and normalize it to the $t\bar{t}gg$ cross section of 4.3 nb. Just as
in the case of $t\bar{t}H$, we assume that the dominant contribution to the
$t\bar{t}$ cross section comes from gluon fusion.  Since the top quarks
are heavy and hardly radiate, we next assume that the two extra jets are
gluons coming from initial state radiation only. The angular and energy
distributions of the gluons, in addition to Eq.~(10), are assumed to follow
the Altarelli-Parisi function
$$
  P_{gg}(z) = c_{gg} \left( \frac{z}{1-z} + \frac{1-z}{z} + z(1-z) \right),
\eqno(11)
$$
where $z$ is the fractional momentum of the initial gluon after radiation:
$$
  z = \frac{E_i - {p_L}_i}{E_{i-1} - {p_L}_{i-1}}~,
\eqno(12)
$$
where the subscripts $i-1$ and $i$ refer to the initial gluon before and
after radiation. We also use the approximation introduced by Dokshitzer et
al.\Ref{13} where the further splitting of a radiated virtual gluon is
replaced by two sequential radiations from the same initial gluon. The $t$
and $\bar{t}$ are decayed as before to give us the required signature.

To mimic a real experimental situation, the energy of each parton is smeared
to reproduce the proposed detector resolution for jets and leptons\Ref{14}.
Then we impose an isolation cut on each of the hadronic partons (jets) in
$\eta$-$\phi$ space so that no two jets are within a distance of 0.7 units
from each other:
$$
  \Delta R = \sqrt{\Delta\eta^2 + \Delta\phi^2} > 0.7~.
$$
A rapidity cut of $\eta<3.0$ is also applied to each of the jets and leptons.

\section{RESULTS}

We generate the number of events for the signal and background that
corresponds to one typical SSC year at the luminosity of
$10^{33}~\rm cm^{-1}s^{-1}$ ($10^4~\rm pb^{-1}$). The following cuts are
then performed on both types of events:
(1) A 20 GeV minimum $p_T$ cut is first applied to all jets and leptons.
(2) A second $p_T$ cut is then applied to the four hadronic jets:
For each of the six pairs of jets, we form the scalar sum of the two
individual $p_T$ and require it to be larger than 80 GeV\null. This will
severely suppresses the background while keeping the signal almost
unchanged.  (3) To try to guarantee that the two leptons originate from top
quark decays, we require each lepton to be at least a distance  $\Delta R =
0.4$ away from any jet in $\eta$-$\phi$ space. The two leptons are required
to be  separated by the same amount. (4) The missing transverse energy in
each event must be larger than 40 GeV.

There are many less important sources of background such as $Z+{\rm 4~jets}
+{\rm missing}~E_T$ and $b\bar{b} + 2g$, etc.  To ensure that they do not
significantly affect our result, we also apply the following cuts:
(1) $Z$-peak cut for leptons: When the lepton pairs are of the same type,
i.e., $e^+e^-$ or $\mu^+\mu^-$, their invariant mass should be outside of
the region $91.2 \pm 5.0$ GeV of the $Z$-peak.
(2) Observed invariant mass: The invariant mass of all the observed
particles should be larger than $2m_t + m_H$, which is 380 GeV in our case.
This requires the average energy of each jet or lepton to be more than 60
GeV in the center of mass, greatly enhancing the probability of having
heavy particles in the final state.

The results of all the cuts are shown in Table~I in the order of their
applications. After these cuts we are left with a signal of 0.071 pb, which
corresponds to roughly 710 events in one SSC year, and a background of 2.8
pb.

$$\vbox{
  \halign{\quad#\hfil & \quad\hfil#\hfil & \quad\hfil#\hfil \cr
  {\bf \hfill Table I.} &&\cr
  \noalign{\vskip1ex\hrule \vskip1pt \hrule \vskip1ex}
	      & Signal ($t\bar{t}H$)      & Background ($t\bar{t}gg$) \cr
  \qquad Cuts & $100\% \approx 0.4$ pb & $100\% \approx 200$ pb \cr
  \noalign{\vskip1ex \hrule \vskip1ex}
  All $p_T > 20$ GeV                      &  54\%  &  66\% \cr
 all  $p_{T1}+p_{T2} > 80$ GeV &  51\%  &  13\% \cr
  Missing $E_T > 40$ GeV                  &  51\%  &  12\% \cr
  Leptonic $Z$-peak cut                   &  41\%  &  9.5\% \cr
  Observed invariant mass $>$ 380 GeV     &  36\%  &  7.2\% \cr
  Isolation cuts ($j$-$j$, $l$-$j$ and $l$-$l$) &  18\%  & ~1.4\% \cr
  \noalign{\vskip1ex \hrule}}
}$$

To extract the signal from this background, we need to reconstruct the
Higgs mass from two-jet pairs. Figure~1a shows the distribution of the
two-jet invariant mass of the signal alone. It has a prominent peak about
$m_H$ = 100 \GeVc2\ containing 650 counts over a combinatorial background
of roughly 130 counts in the 10 \GeVc2\ or so region under the peak.  The
width of the resonance matches roughly the two-jet resolution of 10 GeV
achievable at the SSC.

Figure~1b shows the combined result of both $t\bar{t}H$ and $t\bar{t}Z$.
Next to the Higgs boson peak at 100 GeV is the $Z$-peak at 90 GeV, with
approximately 550 counts above the combinatorial background. We have used a
$t\bar{t}Z$ cross section of 12 pb with 70\% of the $Z$ bosons decaying to
$q\bar{q}$ pairs. This has a combined cross section $\times$ branching ratio
of 8.4 pb, which is very close to the 8 pb we used for $t\bar{t}H$ events.
The slightly lower peak for the $Z$ is due mainly to the two $p_T$ cuts.

The $t\bar{t}gg$ background is shown in Fig.~1c. The shaded area at the
bottom is the result of Fig.~1b plotted on the same scale for comparison.
Over the same 10 GeV range under the Higgs peak, the background contains
roughly 9900 counts.  Thus, we obtain a signal to noise ratio of 650 : 9900
+ 230, for one SSC year, where 100 out of the 230 counts come from the
$t\bar{t}Z$ combinatorial background. This is an 6.5$\sigma$ effect.

The combined result is shown in Fig.~1d.  This figure shows a clear signal
over the QCD background despite the fact that the cuts applied have not been
fully optimized. In view of all the approximations we have made, our
background in the interested region can easily be off by a factor of two or
three. However, we have been rather generous in normalizing the overall
cross section for the background but not for the signal. Together with more
optimal cuts, the signal should still stand out from even a higher than
expected background. We believe that reconstruction from jets is a promising
technique in the detection of the intermediate mass Higgs boson

Before closing, we would like to point out that once $m_t$ is in the region
of 80--200 \GeVc2 the $t\bar{t}H$ cross section does not change appreciably
with the $t$ quark mass.  However, the inclusive $t\bar{t}$ cross section
decreases rapidly from 30 nb at $m_t=100$ \GeVc2 to 0.25 nb at $m_t=200$
\GeVc2 so that a lighter $t$ quark will make reconstruction from jets more
difficult.  On the other hand, the $t\bar{t}H$ cross section decreases from
20 pb to 5 pb\Ref{1}, when $m_H$ increases from 80 to 140 \GeVc2; but the
two-jet invariant mass spectrum of the background decreases even faster, and
it actually favors Higgs bosons of a heavier mass.

The authors wish to thank Tao Han and Frank Paige for many useful
discussions, and Steve Parke and A.~L.~Stange for sharing their computer
codes. D. D. W. would also like to thank the computer staff at the SSC and
physicists at the Prairie View A\&M for assistance, especially David Liu,
Tiantai Song, Jin Chu Wu  and David Wagoner.

\vfil \eject

\section{REFERENCES}

\item{[1]} Z.~Kunszt, Nucl.\ Phys.\ {\bf B247}, 339 (1984);  D.~Dicus and
S.~Willenbrock, Phys.\ Rev.\ D {\bf 39}, 751 (1989).

\item{[2]} W. J.~Marciano and F.~E.~Paige, Phys.\ Rev.\ Lett.\ {\bf 66}, 2433
(1991); J.~Gunion, U. C. Davis preprint UCD-91-2 (1991).

\item{[3]} Z.~Kunszt, Z.~Trocsanyi, and W.~J.~Sterling, Nucl.\ Phys.\
{\bf B271}, 247 (1991); Ren-Yun Zhu, GEM Internal Report, GEM TN-91-32,
November 30, 1991;  H.~Yamamoto
and Ren-Yuan Zhu, GEM Internal Report, GEM TN-92-126, June 1992.

\item{[4]} A. Ballestero and E. Maina, Phys.\ Lett.\ {\bf 268B}, 438 (1991).

\item{[5]} GEM Letter of Intent to the SSC, November 30, 1991.

\item{[6]} D.~Zeppenfeld et al., Univ.\ of Wisconsin preprint MAD-PH-605
(1990); Proc.\ ECFA LHC Workshop in Aachen, Oct.\ 4--9, 1990.

\item{[7]} F.~A.~Berends, J.~B.~Tausk, and W.~T.~Giele, Fermilab-Pub-92/196-T
and references therein.

\item{[8]} See, e.g., F.~J.~Gilman and L.~E.~Price, Proc.\ 1986 Summer Study
on the Physics of the Superconducting Super Collider, Snowmass, edited by
R.~Donaldson and J.~Marx, p.\ 185, and related contributions in the same
volume. These studies focus mainly on the $WH$ channel with a one-lepton
trigger.

\item{[8a]} P. Nason, S. Dawson, and R. K. Ellis, Nucl.\ Phys.\ {\bf B303},
607 (1988);\break
W.~Beenakker et al., Phys.\ Rev.\ D {\bf 40}, 54 (1989).

\item{[9]} V. Barger and A. L. Stange, private communications.

\item{[10]} H.~Baer, X.~Tata, and J.~Woodside, Phys.\ Rev.\ D {\bf 45}, 142
(1992).

\item{[11]} V.~Barger, A.~L.~Stange, R.~J.~N.~Phillips, Phys.\ Rev.\ D
{\bf 45}, 1484 (1992).

\item{[12]} See Ref.~10 and 11. We infer from their $W/Z$ ratio that
$t\bar{t}Z$ is also dominated by gluon fusion.

\item{\rm [13]} Y.~L.~Dokshitzer, V.~A.~Khoze, and S.~A.~Troyan in {\it
Advanced Series on Directions in High Energy Physics, Vol.\ 5}, edited by
A.~H.~Mueller (World Scientific, Singapore, 1988) p.\ 241.

\item{[14]} SDC Letter of Intent to the SSC, SSC-EOI-0020 (1991),
and Ref.\ 5.

\vfil\eject

\section{FIGURE CAPTIONS}

FIG.~1. Two-jet invariant mass distributions. The vertical axes shows
counts/2 GeV bin. The horizontal axes shows two-jet invariant mass
values in GeV/$c^2$.
(a) Signal, $t\bar{t}H$ events.
(b) $t\bar{t}H$ and $t\bar{t}Z$ events.
(c) QCD background $t\bar{t}gg$ events (open histogram), and signal from (b)
$t\bar{t}H + t\bar{t}Z$ (shaded histogram).
(d) Signal plus background. The shaded areas
represent $t\bar{t}H + t\bar{t}Z$ events from (b).

\vfil\eject

\bye